# Challenges in orbital current-driven domain wall motion in light metal/ferrimagnet heterostructures

Min-Gu Kang[1,3]*, Jaerin Kim[1,3], Benjamin J. Jacot[1], Laura Van Schie[1,2], and Pietro Gambardella[1]*

[1]*Department of Materials, ETH Zurich, 8093 Zurich, Switzerland*

[2]*Department of Physics, ETH Zurich, 8093 Zurich, Switzerland*

[3]*These authors equally contributed to this work*

* Correspondence to: kang.mingu.d7@tohoku.ac.jp (M.-G.K) & pietro.gambardella@mat.ethz.ch (P.G)

## ABSTRACT

**Recent advances in spintronics suggest that orbital Hall currents generated by charge injection in light metals can provide nonequilibrium angular momentum without relying on strong spin–orbit coupling (SOC). Here, we examine whether such orbital currents can drive domain wall (DW) motion in an amorphous ferrimagnetic alloy, $Gd_{25}(Fe_9Co_1)_{75}$ (GFC), where the rare-earth sublattice offers strong SOC for orbital-to-spin conversion. We compare three representative heterostructures: Pt/GFC as a spin Hall reference, light-metal (Mn or Ti)/GFC for direct orbital-current injection, and light-metal/Pt/GFC incorporating an ultrathin Pt layer for orbital-to-spin conversion. Whereas Pt/GFC exhibits robust and reproducible spin–orbit-torque-driven DW motion, no current-driven DW motion is detected in Mn/GFC or Ti/GFC. Second-harmonic Hall measurements nevertheless reveal finite damping-like torques in both Mn/GFC and Ti/GFC, demonstrating that angular-momentum transfer into GFC does occur but is far weaker than in Pt/GFC. Inserting a 1-nm-thick Pt conversion layer strongly enhances the damping-like torque and restores DW motion. Thickness-dependent analysis further shows that DW mobility and depinning thresholds correlate with the interfacial Dzyaloshinskii–Moriya interaction and domain-wall width, highlighting weak torque conversion and insufficient interfacial stabilization of chiral DWs as key challenges for orbital-driven DW motion in light-metal/ferrimagnet heterostructures.**

## I. INTRODUCTION

The generation and manipulation of spin currents play a central role in spintronics, enabling efficient electrical control of magnetization through spin transfer torques[1–4] and spin–orbit torques (SOTs)[5–7]. In conventional heavy-metal/ferromagnet heterostructures, strong spin–orbit coupling (SOC) in heavy elements such as Pt, Ta, and W gives rise to the spin Hall effect (SHE)[8], which converts charge currents into transverse spin currents that can exert torques on adjacent magnetic layers[6,9–11]. This mechanism has been successfully exploited to drive magnetic switching[5,12–16], domain wall (DW) motion[17–20], and skyrmion dynamics[21–25], forming the basis of a wide range of spintronic devices[3,6,26,27].

Recently, theoretical work suggested alternative pathways for generating angular momentum based on the orbital Hall effect (OHE), where a transverse flow of orbital angular momentum is generated from a longitudinal charge current even in materials with weak SOC[26–28]. Because orbital currents do not rely on strong SOC for their generation, light metals such as Mn and Ti have been predicted and experimentally suggested to exhibit sizable orbital Hall conductivities[29–33]. If efficiently converted into spin angular momentum at magnetic interfaces, orbital currents could provide a new route for torque generation without using heavy metals, offering potential advantages for energy-efficient spintronic devices[31,34–36].

Open questions in the field concern the definition of orbital currents and orbital transport[37–39], as well as the interplay of alternative effects generating local accumulation of orbital momenta[40,41]. Another key unresolved question is whether orbital currents (or orbital accumulation) can effectively drive DW dynamics in magnetic heterostructures. While several studies have reported current-induced magnetic oscillations[31,33–35,42] and switching[36,43–45] due to orbital torques, direct experimental evidence of orbital-driven DW motion in magnetic materials remains scarce. In particular, it remains unclear whether orbital currents injected into ferrimagnetic alloys, which are well-known for their fast DW dynamics[46–52], can generate sufficient torque to overcome pinning and drive DW motion.

In this work, we address this question by investigating current-induced DW motion in Gd-rich ferrimagnetic $\mathrm{Gd_{25}(Fe_9Co_1)_{75}}$ (GFC) thin films with perpendicular magnetic anisotropy (PMA). The central goal is to test in which conditions an orbital current generated in a light metal can, upon injection into a rare-earth-containing ferrimagnet, produce a measurable current-induced torque that is sufficient to drive DW motion. GFC provide a particularly relevant platform for this test because of its PMA, fast SOT-driven DW dynamics in ferrimagnets, and different orbital-to-spin angular momentum conversion channels afforded by Gd and FeCo[31,42].

The experimental concept is summarized in **Figs. 1(a,b)**. In a light metal (LM)/GFC bilayer [**Fig. 1(a)**], a longitudinal charge current in the LM layer (Mn or Ti) is expected to generate a transverse orbital accumulation. Once injected into GFC, this orbital angular momentum can be converted into spin angular momentum through local atomic SOC[42]. To explicitly reflect this process, we introduce sublattice-resolved orbital-to-spin conversion efficiencies, $\eta_{LS}^{\mathrm{Gd}}$ and $\eta_{LS}^{\mathrm{FeCo}}$, which quantify the effectiveness of L·S coupling in the Gd (5d) and FeCo (3d) sublattices, respectively. Because the signs of the SOC on the two sublattices are opposite, the spin polarizations generated via $\eta_{LS}^{\mathrm{Gd}}$ and $\eta_{LS}^{\mathrm{FeCo}}$ are expected to have opposite signs[42]. As a result, even if a finite orbital current is injected into GFC, the net effective torque acting on the DW can be strongly reduced by partial cancellation between the two sublattice contributions. In the LM/Pt(1)/GFC trilayer [**Fig. 1(b)**], we insert an ultrathin Pt spacer to introduce an additional orbital-to-spin conversion channel characterized by $\eta_{LS}^{\mathrm{Pt}}$. In this structure, the orbital current generated in the LM layer can be converted into a spin current inside the Pt layer, where the conversion efficiency and spin polarization are mainly governed by Pt's strong SOC. Pt can also generate a spin current directly via the conventional spin Hall effect, but this effect is considered to be small for Pt layers thinner the spin diffusion length[34]. These spin currents are injected into GFC and exert torques on the DW, providing a controlled comparison to the direct-injection case in LM/GFC.

By directly comparing Pt/GFC, LM/GFC, and LM/Pt(1)/GFC structures, we systematically examine how the efficiency of orbital-to-spin conversion at different interfaces, parameterized by $\eta_{LS}$, influences the magnitude of current-induced torque and the resulting DW dynamics. Combined with real-space polar magneto-optical Kerr effect (MOKE) imaging and quantitative second-harmonic Hall measurements, this approach allows us to establish the practical requirements and limitations for realizing orbital-current-driven DW motion in light-metal/ferrimagnet heterostructures.

## II. EXPERIMENTAL METHODS

Thin film heterostructures were deposited on thermally oxidized Si substrates by magnetron sputtering at room temperature with a base pressure below $3.0 \times 10^{-8}$Torr. All metallic layers were deposited by DC sputtering with a working pressure of 3 mTorr. The ferrimagnetic layer consists of GFC with a nominal composition of $\mathrm{Gd_{25}(Fe_9Co_1)_{75}}$, chosen to maintain strong PMA and achieve a Gd-dominant ferrimagnetic state. The Gd concentration was controlled by adjusting the deposition rates of $\mathrm{Fe_9Co_1}$ and Gd using a confocal sputtering configuration. The GFC thickness was fixed at 15 nm for all samples. To deposit the LM/Pt/GFC trilayers, a 1-nm-thick Pt spacer layer was sputtered sequentially to the LM layer without breaking vacuum to ensure clean interfaces. The light-metal thickness $t_{\mathrm{LM}}$ was varied from 3 to 17 nm. All samples were capped with 8 nm of SiN to prevent surface oxidation of GFC.

The magnetic properties of the GFC layer were characterized using a superconducting quantum interference device (SQUID) magnetometer. Out-of-plane hysteresis loops were measured at room temperature under magnetic fields up to 3 T to extract the saturation magnetization $M_s$ and coercive field. **Figure 1(c)** shows a representative hysteresis loop measured for a Mn(12)/GFC(15) sample (thickness in nanometers), exhibiting square switching behavior consistent with strong PMA. To characterize the ferrimagnetic state, the anomalous Hall resistance $\mathrm{R_H}$ was measured using a standard four-point probe geometry. As shown in **Fig. 1(d)**, the Hall signal shows a negative sign at room temperature, consistent with a Gd-rich ferrimagnetic configuration[52,72–74]. To further verify the magnetic compensation behavior, temperature-dependent magnetization measurements were performed using SQUID (**Fig. S3**). The net magnetization does not exhibit a compensation point within the measured temperature range up to at least 400 K, indicating that the system remains in the Gd-dominant regime over the entire experimental range. Accordingly, the measurements reported in this work are conducted within the Gd-dominant regime.

Racetrack-shaped microdevices were fabricated using standard photolithography and lift-off processes. The resulting devices consist of 10-μm-wide and 60-μm-long tracks with electrical contact pads at both ends. Current-induced DW motion was investigated using polar MOKE microscopy. A DW was nucleated by applying an out-of-plane magnetic field pulse. Following DW nucleation, a train of $N$ = 10 identical current pulses was applied along the racetrack under each measurement condition to drive DW motion. The total longitudinal DW displacement $\Delta x$ accumulated over the pulse train was determined from differential MOKE images acquired before and after pulse injection. As illustrated in **Fig. 1(e)**, DWs were nucleated and prepared in a well-defined configuration (Up–Down or Down–Up) prior to current pulse application. Unipolar or bipolar current pulses with amplitudes of 10–45 V and pulse durations of 10–100 ns were applied through the device contact pads. DW displacement was visualized using differential Kerr imaging, which subtracts a reference frame to highlight net DW motion with high contrast. An example differential image is shown in **Fig. 1(f)**. The average DW velocity was defined operationally as $v_{\mathrm{DW}} = \Delta \mathrm{x}/(N t_{\mathrm{p}})$, where $t_{\mathrm{p}}$ is the nominal width of each pulse and $N$ = 10. Thus, the reported velocity corresponds to the total DW displacement accumulated over ten identical pulses divided by the total nominal pulse-on time. The same pulse protocol was used consistently for all samples. The measured voltage waveform exhibited a rise time $t_{\mathrm{rise}} \approx 2$ ns, defined by the conventional 10-90% criterion. The reported $v_{\mathrm{DW}}$ values are therefore accurate up to $2t_{\mathrm{rise}}/t_{\mathrm{p}} \leq 10\%$. Current densities were defined as $J_{\mathrm{Pt}}$, $J_{\mathrm{LM}}$, and $J_{\mathrm{Pt+LM}}$ for currents flowing in the Pt layer, light-metal layer, and LM/Pt bilayer, respectively, taking into account the layer resistivities and thicknesses. Current density values were calculated using a parallel resistor model (see **Supporting Information S2** for details). All measurements were performed at room temperature under ambient conditions.

## III. RESULTS AND DISCUSSION

### Current-driven DW motion in Pt/GdFeCo bilayers

To establish a reliable baseline and validate our experimental approach, we first investigate current-induced DW dynamics in Pt(t)/GFC(15 nm) bilayers. Pt is a prototypical spin Hall material that generates transverse spin currents under an in-plane charge current via the SHE, thereby exerting SOTs on adjacent magnetic layers[53–57]. This makes Pt/GFC an ideal reference system for comparison with orbital-driven DW measurements.

**Figure 2(a)** schematically illustrates the Pt(5)/GFC(15) sample used for DW motion measurements. The ferrimagnetic GFC layer consists of two antiferromagnetically coupled sublattices: the rare-earth (RE) sublattice dominated by Gd (red) and the transition-metal (TM) sublattice (blue). A Down-to-Up DW is initialized, separating a left-side down-magnetized domain ($-M_z$) from a right-side up-magnetized domain ($+M_z$). When a longitudinal current is applied to the Pt layer, a transverse spin current is generated via the SHE and injected into the GFC layer across the interface[55–57]. This spin current exerts a damping-like SOT that can effectively drive the propagation of a Néel-type DW, provided it is stabilized by interfacial Dzyaloshinskii–Moriya interaction (DMI)[48,58,59].

A representative differential Kerr image acquired after a current pulse in shown in **Fig. 2(b)**. The parallelogram-shaped contrast indicates the net displacement of the DW along the racetrack. Under a positive current density ($+J_{Pt}$), the DW moves in the $+x$ direction, whereas reversing the current polarity ($-J_{Pt}$) reverses the direction of motion. This deterministic polarity-dependent behavior is a hallmark of SOT-driven dynamics[20,47,48], and cannot be explained by purely thermal or field-driven mechanisms.

We systematically measured $v_{DW}$ as a function of the applied current density in the Pt layer, $J_{Pt}$, for Pt thickness of 5 nm. Representative results for Pt(5)/GFC(15) are shown in **Fig. 2(c)**, where the pulse length was fixed at $t_p = 20$ ns. Beyond the depinning regime, three characteristic regions are observed: (i) a linear increase of $v_{DW}$ in the flow regime ($0.2 – 0.4 \times 10^{12}$ A/m$^2$), (ii) a velocity saturation region ($0.5 – 0.8 \times 10^{12}$ A/m$^2$), and (iii) a second rise beyond $0.8 \times 10^{12}$ A/m$^2$, which we attribute to Joule heating at high current density (see **Supporting Information S3**). This behavior is consistent with previous studies on Pt/RE-TM ferrimagnetic systems, such as Pt(6)/$Gd_{44}Co_{56}$ and Pt(4)/$Co_{1-x}Gd_x$, where velocities of 350 – 900 m/s have been reported depending on temperature and composition[48,59].

We define the effective DW mobility as $\mu_{DW} = \frac{dv_{DW}}{dJ_{Pt}}$ evaluated in the low-current linear flow regime ($0.2 \times 10^{12}$ A/m$^2 < J_{Pt} < 0.4 \times 10^{12}$ A/m$^2$) where Joule heating effects are less pronounced. Although thermal effects can never be completely eliminated in measurements of current-induced DW motion, $\mu_{DW}$ provides a consistent comparison of current-driven DW dynamics among the different heterostructures measured in this work using the same pulse protocol. Furthermore, $\mu_{DW}$ is defined consistently with other studies of current-induced DW motion in ferrimagnetic systems[75,76]. For Pt(5)/GFC(15), we obtain $\mu_{DW} = (6.0 \pm 0.7) \times 10^{-10}$ m$^3$/As, which is significantly higher than that reported for typical ferromagnetic systems such as Pt/CoFe/MgO (~$0.4 \times 10^{-10}$ m$^3$/As)[17] and is comparable to values reported for Pt/$Gd_{44}Co_{56}$ (~$10 \times 10^{-10}$ m$^3$/As)[48]. These comparisons highlight the high efficiency of SOTs in driving DW motion in GFC.

To probe the role of DMI and identify the DW chirality, we measured the dependence of $v_{DW}$ on an in-plane magnetic field $H_x$, applied along the racetrack direction[17,18,60]. As shown in **Fig. 2(d)**, for Down-to-Up DWs, a positive $H_x$ enhances $v_{DW}$, whereas a negative $H_x$ suppresses $v_{DW}$. This asymmetry is reversed for Up-to-Down DWs, indicating a left-handed DW chirality consistent with a negative effective DMI field. Such behavior agrees with earlier reports on Pt/GdCo and Pt/CoTb systems[47,48,59], where interfacial DMI stabilizes left-handed Néel-type DWs that couple efficiently to SOTs.

Overall, these results confirm that Pt/GFC bilayer structures reliably support current-induced SOT-driven DW motion. The quantitative agreement with established benchmarks validates our measurement methodology and establishes Pt/GFC as a robust reference platform for subsequent investigation of orbital-current-driven DW dynamics in light-metal-based structures.

**Absence of current-driven DW motion in Mn/GdFeCo and Ti/GdFeCo bilayers**

Having established efficient spin current-driven DW motion in Pt/GFC bilayers, we now turn to the central question of this study: can orbital currents generated in LMs such as Mn or Ti drive DW motion in a ferrimagnetic layer?

To address this question, we investigate a series of LM($t_{LM}$)/GFC(15) bilayers with LM = Mn or Ti, where $t_{LM} = 3$ to 17 nm. Recent theoretical and experimental studies have suggested that these LMs possess sizable orbital Hall conductivities, enabling the generation of transverse orbital currents from longitudinal charge currents without relying on strong SOC[10–12,19,20]. Based on this premise and the expected orbital-to-spin conversion at the LM/GFC interface and inside of GFC, we expected to observe at least a partial DW response under applied current pulses.

Contrary to this expectation, we did not observe any reliable or directional DW motion in either Mn/GFC or Ti/GFC bilayers. As shown in **Figs. 3(a,b)**, despite systematic variation of the current density $J_{LM}$ and pulse duration $t_p$, no reliable or directional DW motion was observed in Mn/GFC or Ti/GFC bilayers. In the regime of moderate current injection (e.g. $J_{LM} \sim 0.05 - 0.15 \times 10^{12} \mathrm{A/m^2}$, $t_p$=20–40 ns), no discernible change in Kerr contrast was detected, indicating that DWs remained pinned. To approach a possible depinning regime, we further increased the current density up to $J_{LM} \sim 0.25 \times 10^{12} \mathrm{A/m^2}$ and extended the pulse duration up to $t_p$= 100 ns, approaching the thermal threshold for random nucleation[34,35]. Under these extreme conditions, isolated and stochastic domain nucleation events were occasionally observed [**Fig. 3(b)**], which we attribute to Joule heating. Importantly, even in high-current regime, reproducible directional DW motion was not achieved.

To further probe the possible presence of weak orbital torque effects, we performed in-plane magnetic field-assisted DW motion measurements on representative Mn($t_{LM}$)/GFC(15) samples ($t_{LM}$ = 5 and 12 nm), as shown in **Figs. 3(c-d)**. Here, the DW displacement normalized by the pulse length is plotted as a function of the in-plane field $H_x$ for fixed current densities ($J_{LM} = 0.23 \times 10^{12} \mathrm{A/m^2}$ for Mn(12) and $0.32 \times 10^{12} \mathrm{A/m^2}$ for Mn(5)). If a finite current-induced torque were present, an in-plane magnetic field $H_x$ would stabilize a Néel-type DW configurations and thereby systematically influence the DW mobility by tuning the coupling efficiency between the torque and the DW[29]. However, no clear or reproducible dependence on $H_x$ was observed. The small apparent displacements exhibited large error bars, lacked directionality, and varied randomly between measurements, indicating that they are most likely arising from thermal activation, field-dependent spin transfer torque in the creep regime[77], or instrumental noise. Notably, all measured signals remained well below the threshold used in the Pt/GFC reference to define reliable DW motion.

Complementary SQUID magnetometry and anomalous Hall measurements confirmed that all LM($t_{LM}$)/GFC(15) samples exhibit robust PMA, with saturation magnetization $M_s$ and coercive fields comparable to those of Pt/GFC. This confirms that the intrinsic magnetic properties of the GFC layer are not responsible for the absence of DW motion.

Taken together, these results demonstrate that orbital Hall accumulation generated in Mn or Ti do not produce sufficient torque to depin and drive DWs in GFC under our experimental conditions. This indicates that one or more of the key processes required for orbital-current-driven DW dynamics[19,20], namely (i) orbital current generation in the LM layer, (ii) transmission across the LM/GFC interface, (iii) orbital-to-spin conversion within GFC, and (iv) DMI stabilization of Néel DWs are insufficient or strongly suppressed in these bilayer structures. In addition to the weak torque magnitude resulting from (i-iii), the internal DW configuration may also play an important role. In systems with weak interfacial spin–orbit coupling such as LM/GFC bilayers, the interfacial DMI is expected to be vanishing, which favors Bloch-type DWs rather than chiral Néel-type DWs. Because damping-like torques couple most efficiently to Néel DWs[60], the lack of chiral stabilization would further reduce the efficiency of current-driven DW propagation. While our measurements of field-assisted DW motion in Figs. 3(c,d) suggest that the torque magnitude itself is extremely small, the additional role of DW structure cannot be excluded at this stage.

We thus consider two primary microscopic scenarios underlying the weak net torque. First, orbital current injection may be hindered by interfacial disorder, such as atomic intermixing, roughness, or partial oxidation, which could disrupt coherent orbital overlap at the LM/GFC interface[61,62]. Second, even if orbital angular momentum is successfully injected into GFC, the subsequent orbital-to-spin conversion may lead to a strongly reduced net torque due to the ferrimagnetic sublattice structure of GFC. Because the SOC in the Gd and FeCo sublattices has opposite signs, the spin polarizations generated via orbital-to-spin conversion on each sublattice are expected to compete with each other[31,42], resulting in partial cancellation of the effective torque acting on the domain wall. To distinguish between these possibilities and to directly assess whether a finite current-induced torque exists in LM/GFC, we evaluate next the torque magnitude using second-harmonic Hall measurements.

**Quantitative evaluation of current-induced torque**

In Fig. 3, we showed that no current-driven DW motion is observed in LM/GFC bilayers. To clarify whether this absence originates from a complete lack of current-induced torque or from an insufficient torque magnitude, we directly quantify the torque using second-harmonic Hall measurements, a well-established technique for evaluating spin–orbit torques in perpendicularly magnetized systems[6,10,63].

**Figure 4(a)** shows representative first- ($R_{xy}^{1\omega}$) and second-harmonic ($R_{xy}^{2\omega}$) Hall resistsance measured as a function of an in-plane magnetic field $B_x$ for a Pt(5)/GFC(15) device. For a sample with PMA, the first-harmonic Hall resistance can be expressed as $R_{xy}^{1\omega} = R_{AHE}\cos\theta + R_{PHE}\sin^2\theta\sin 2\phi$, where $R_{AHE}$ and $R_{PHE}$ are the anomalous and planar Hall resistances, respectively, and $\theta$ and $\phi$ denote the polar and azimuthal angles of the magnetization[10,63]. Under a small in-plane field $B_x$, the magnetization is slightly tilted away from the out-of-plane easy axis, resulting in a parabolic dependence of $R_{xy}^{1\omega}$ on $B_x$. Accordingly, the low-field region of $R_{xy}^{1\omega}$ is fitted with a quadratic function, as indicated by the cyan curve in **Fig. 4(a)** from which the curvature $C$ is extracted. For completeness, representative first- and second-harmonic Hall signals measured for Mn(8)/GFC, Mn(8)/Pt(1)/GFC, Ti(8)/GFC, and Ti(8)/Pt(1)/GFC devices, together with their corresponding low-field fitting procedures, are summarized in **Supporting Information S4**.

The $R_{xy}^{2\omega}$ originated from the oscillation of the magnetization induced by the AC current through current-induced effective fields. In the low-field regime ($B_x \ll B_{eff}$), $R_{xy}^{2\omega}$ varies linearly with $B_x$, and can be written as $R_{xy}^{2\omega} = \frac{\partial R_{xy}}{\partial\theta}\frac{B_{DL}}{B_{eff}}B_x$, where $B_{DL}$ is the damping-like effective field generated by the applied current, and $B_{eff}$ is the effective field[63], dominated by the perpendicular magnetic anisotropy and demagnetization fields. The linear fitting of $R_{xy}^{2\omega}$ in the low-field region is shown by the yellow line in **Fig. 4(a)**, from which the slope $S$ is obtained. Following the standard analysis, the damping-like effective field is extracted as $B_{DL} = -\frac{2S}{C}$. To compare different heterostructures on an equal footing, we normalize $B_{DL}$ by the applied electric field $E$, yielding the damping-like Hall conductivity $\xi_{DL}^{E} = \frac{2e}{\hbar}\frac{M_s t_{GFC}}{E}B_{DL}$, where $M_s$ is the saturation magnetization and $t_{GFC}$ is the thickness of the GFC layer[6]. This quantity directly reflects the efficiency of angular-momentum transfer per unit electric field and allows quantitative comparison among different samples.

**Figure 4(d)** summarizes the extracted $\xi_{DL}^{E}$ (in units of $10^5\Omega^{-1}m^{-1}$) for the different heterostructures. For the Pt(5)/GFC reference sample, we obtain a large positive value of $\xi_{DL}^{E} = +8.3 \pm 0.5$, which is consistent with the well-know positive spin Hall angle of Pt and confirms that spin currents generated in Pt effectively exert a damping-like torque on GFC[64,65].

In contrast, Mn(8)/GFC and Ti(8)/GFC bilayers exhibit much smaller magnitudes of $\xi_{DL}^{E}$, with values of $-0.15\pm0.05$ and $-0.05\pm0.03$, respectively. Despite their small magnitudes, these finite values demonstrate that angular momentum transfer from the light metals into GFC does occur. Notably, both values have a negative sign, opposite to that observed in Pt/GFC. This sign reversal provides important physical insight. In the LM/GFC structures, a positive orbital current generated in the light metal is injected into GFC and converted into a spin current primarily through the Gd sublattice. Because the spin–orbit coupling of the Gd 5d states has a negative sign, the resulting spin polarization, and thus the damping-like torque, naturally acquires a negative sign. This observation is fully consistent with the schematic picture introduced in **Fig. 1(a)**. The gray dashed line in **Fig. 4(d)** corresponds to the value measured in a single GFC layer, $\xi_{\mathrm{DL}}^{E} = +0.01 \pm 0.01$, which is nearly negligible. This small positive background likely originates from self-torque effects intrinsic to GFC and sets the baseline for evaluating interfacial torque contributions. The fact that the values measured in Mn/GFC and Ti/GFC are only slightly larger than this background further highlights the weak net torque generated by direct orbital-current injection.

The small net torque observed in LM/GFC can arise from several, not mutually exclusive, mechanisms. First, the orbital Hall conductivity of Mn and Ti may be intrinsically small, resulting in weak orbital-current generation, consistent with previous reports on Mn-based heterostructures[31]. Second, even if an orbital current is generated, its conversion into spin angular momentum within GFC may be inefficient due to the antiparallel alignment the opposite signs of the L·S coupling in the Gd 5d and FeCo 3d states, which can lead to partial cancellation of the resulting spin currents. Third, contributions from self-torque effects intrinsic to GFC cannot be excluded and may further reduce the effective torque acting on the DWs.

Having established that a finite but insufficient torque exists in LM/GFC, we now show that inserting an ultrathin Pt layer fundamentally alters this situation. In the LM/Pt(1)/GFC structures, $\xi_{\mathrm{DL}}^{E}$ increases dramatically to +1.6 $\pm$0.4 for Mn(8)/Pt(1)/GFC and +1.0 $\pm$0.3 for Ti(8)/Pt(1)/GFC. Importantly, the sign of $\xi_{\mathrm{DL}}^{E}$ reverses compared to the corresponding LM/GFC bilayers and becomes positive, matching the sign observed in Pt/GFC. In the LM/Pt(1)/GFC geometry, the orbital current generated in the light metal is efficiently converted into spin angular momentum inside the 1-nm-thick Pt layer, where both the magnitude and sign of the resulting torque are governed by Pt's strong SOC. In the LM/Pt(1)/GFC structures, sputtered Pt with a thickness of 1 nm can still provide a finite conventional spin Hall contribution, and the Pt interlayer should therefore not be regarded as a negligible spin-current source. However, the observed enhancement is not well described by a simple picture in which Pt(1) contributes only an additional thickness-scaled SHE torque. Instead, the ultrathin Pt layer should be regarded as a spin–orbit-active interfacial layer that can simultaneously provide a finite SHE contribution and modify the interfacial conditions for angular-momentum transfer and domain-wall stabilization, including the effective DMI. Within this framework, the insertion of Pt(1) qualitatively changes both the torque generation and the domain-wall dynamics relative to the direct LM/GFC bilayers.

**Recovery of current-driven DW motion in Mn/Pt/GdFeCo and Ti/Pt/GdFeCo trilayers**

In addition to the markedly enhanced orbital torque in LM/Pt(1)/GFC, the insertion of a Pt(1) layer with strong SOC is also expected to induce a finite interfacial DMI at the Pt/GFC interface[54–57]. The emergence of such interfacial DMI would stabilize chiral Néel-type domain walls, thereby providing favorable conditions for efficient current-driven DW motion. Motivated by this expectation, we next directly investigate current-induced DW dynamics in LM/Pt(1)/GFC trilayer structures.

**Figures 5(a)** and **5(b)** show the DW velocity $v_{\mathrm{DW}}$ as a function of the effective current density flowing in the nonmagnetic layers, $J_{\mathrm{Pt+LM}}$, for Mn($t_{\mathrm{LM}}$)/Pt(1)/GFC(15) and Ti($t_{\mathrm{LM}}$)/Pt(1)/GFC(15), respectively. For the Mn-based devices, the LM thicknesses are $t_{\mathrm{LM}}$ = 3, 8, 12, and 17 nm, while for the Ti-based devices $t_{\mathrm{LM}}$ = 5, 8, 12, and 17 nm. In stark contrast to the LM/GFC bilayers (Fig. 3), all trilayer samples exhibit robust and reproducible current-driven DW motion. The direction of DW propagation reverses when the current polarity is reversed, confirming that the motion is driven by current-induced torques rather than thermal effects[47]. This result is consistent with DW motion induced by the enhanced orbital torques reported for LM/Pt(1)/GFC trilayers in the previous section.

Additional contributions due to spin transfer torque (STT) may be present, as the nonadiabatic STT is known to induce DW motion in single-layer GdFeCo films in proximity of the angular momentum compensation temperature[78]. Our GdFeCo layers, however, have a magnetic compensation temperature $T_M$ larger than 400 K (Fig. S3). As the angular momentum compensation temperature scales with $T_M$, and is typically larger than $T_M$ in GdFeCo[79], we consider it unlikely that a nonadiabatic STT contributes significantly to DW motion in the present case. Moreover, the combined analysis of current partitioning and control measurements indicates that STT alone cannot account for the observed heterostructure-dependent behavior. In particular, a substantial fraction of the applied current flows through the GdFeCo layer in LM/GFC bilayers, yet no deterministic domain wall motion is observed under these conditions, as shown in Figs. 3(c-d). Furthermore, control measurements on GdFeCo single-layer devices show only weak, non-directional displacements without clear signatures of robust torque-driven motion (**Supporting Information S5**). Taken together, these results and considerations indicate that the strong recovery of deterministic and polarity-dependent domain wall motion in LM/Pt(1)/GFC is mostly due to the enhanced damping-like torque and interfacial spin-orbit effects due to the insertion of the Pt spacer.

To further characterize the nature of the DWs and quantify the interfacial DMI, we measured the DW velocity as a function of an in-plane magnetic field $H_x$, as shown in **Figs. 5(c)** and **5(d)**. All measurements were performed in the Down-to-Up DW configuration with a fixed pulse length of 20 ns. For both Mn/Pt/GFC and Ti/Pt/GFC devices, $v_{\mathrm{DW}}$ exhibits a pronounced asymmetric dependence on $H_x$: positive fields enhance the DW velocity, whereas negative fields suppress it. This characteristic asymmetry is a hallmark of chiral Néel-type DWs stabilized by interfacial DMI[17]. Following established models for SOT-driven DW motion, the effective DMI field $H_{DMI}$ is extracted from the in-plane field at which $v_{\mathrm{DW}}$ reaches a minimum in the $v_{\mathrm{DW}}(H_{\mathrm{x}})$ curve. This condition corresponds to the compensation of the internal $H_{DMI}$ by the external in-plane field, reducing the Néel character of the DW and thereby minimizing the efficiency of SOT-driven propagation. From the extracted $H_{DMI}$, the effective DMI constant is calculated using $D = \mu_0 M_s \Delta_{DW} H_{DMI}$, where $\Delta_{DW}$ is the domain wall width. To further assess the stability of Néel-type domain walls under the present conditions, we estimated the domain-wall anisotropy field $\mathrm{H_k}$ using $\mathrm{H_k} \approx \mathrm{M_s t} \ln(2)/(\pi\Delta_{\mathrm{DW}})$ (**Supporting Information S6**). The estimated $\mathrm{H_k}$ is substantially smaller than the extracted $\mathrm{H_{DMI}}$, supporting the stabilization of Néel-type domain walls in the absence of an applied in-plane field. These results demonstrate that the insertion of a Pt(1) layer not only restores current-driven DW motion but also enables a full quantitative characterization of DW dynamics, including the DW mobility and interfacial DMI strength.

To gain further insight into the mechanisms governing the recovered current-driven-DW motion in LM($t_{LM}$)/Pt(1)/GFC trilayers, we systematically analyze the dependence of key dynamical parameters on the light-metal thickness $t_{LM}$, as summarized in **Fig. 6**. For all devices, $v_{\mathrm{DW}}$ initially increases linearly with $J_{\mathrm{Pt+LM}}$ above the depinning threshold, indicating a flow regime where viscous DW motion dominates[47,67–69]. **Figure 6(a)** shows the DW mobility $\mu_{DW} = \frac{\mathrm{d}v_{\mathrm{DW}}}{\mathrm{d}J_{\mathrm{Pt+LM}}}$ extracted from the linear flow regime of the $v_{DW}(J_{Pt+LM})$ curves (Fig. 5) as a function of $t_{LM}$. Although finite rise and fall times of the current pulses, as well as thermal contributions, may introduce residual contributions to $\mu_{DW}$, the former are expected to be relatively small under the pulse conditions employed here ($\mathrm{t_p} \gg \mathrm{t_{rise}}$). Within these limitations, this procedure provides a consistent basis for comparing relative trends in DW dynamics across different trilayer structures.

For the Mn/Pt(1)/GFC series, $\mu_{DW}$ increases with increasing Mn thickness, reaching a maximum at $t_{LM}$ = 12 nm, and then decreases at 17 nm. In contrast, the Ti/Pt(1)/GFC devices exhibit the highest mobility at the smallest thickness ($t_{LM}$ = 5 nm), followed by a monotonic decrease with increasing Ti thickness. Notably, a pronounced drop in mobility is observed between 8 and 12 nm. Importantly, these nonmonotonic and material-dependent trends cannot be explained by a simple drift–diffusion picture of either spin Hall or orbital Hall currents. If bulk spin or orbital current generation were the dominant factor, one would expect $\mu_{DW}$ to increase with $t_{LM}$ and eventually saturate due to spin/orbital current relaxation[31–36,70,71]. The experimentally observed behavior therefore indicates that the DW dynamics in these trilayers are not governed by the magnitude of the injected angular momentum current alone.

To identify the physical origin of the mobility variations, we analyze different magnetic parameters that determine the DW dynamics. **Fig. 6(b)** summarizes the effective DMI constant $D$, extracted from the in-plane field dependence of $v_{DW}$(Fig. 5). For Mn/Pt(1)/GFC, $D$ exhibits a strong thickness dependence and reaches a maximum at $t_{LM} = 12$nm, precisely matching the thickness at which the DW mobility is maximized. This correlation strongly suggests that $\mu_{DW}$ enhancement in Mn-based trilayers is primarily due to the strengthening of interfacial DMI, which stabilizes chiral Néel-type

DWs and enhances their coupling to damping-like torques. In contrast, the Ti/Pt(1)/GFC series shows a comparatively weaker variation in $D$ with thickness, which does not directly mirror the pronounced drop in $\mu_{DW}$ observed between 8 and 12 nm. This indicates that DMI alone cannot account for the thickness-dependent DW dynamics in the Ti-based devices.

To further elucidate this behavior, we examine $H_{DMI}$ and $\Delta_{DW}$, shown in **Figs. 6(c)** and **6(d)**, respectively. $\Delta_{DW}$ is estimated as $\Delta_{DW} = \sqrt{A_{\mathrm{ex}}/K_{\mathrm{eff}}}$, where $A_{\mathrm{ex}}$ is exchange stiffness (assumed ~7 pJ/m)[48,72], and $K_{\mathrm{eff}}$ is the effective anisotropy energy per volume, $K_{\mathrm{eff}} = \frac{1}{2}\mu_0 M_s H_{\mathrm{ani}}$, obtained from angle-dependent magnetotransport measurements (see details in **Supporting Information S7**). $M_s$ is determined via SQUID magnetometry.

While $H_{DMI}$ follows trends similar to $D$, $\Delta_{DW}$ exhibits a strong LM dependence, particularly in the Ti/Pt(1)/GFC series. Notably, $\Delta_{DW}$ decreases sharply between $t_{LM} = 8$ and 12 nm, coinciding with the abrupt reduction in $\mu_{DW}$ observed in Fig. 6(a). The observed $\mu_{DW}$ cannot be attributed solely to the damping-like torque. Instead, it is governed by the internal DW structure, including the domain-wall width and DMI-stabilized Néel-type configuration. The estimated DW width $\Delta_{\mathrm{DW}}$, derived from $\mathrm{K_{eff}}$, shows a trend consistent with the measured mobility, indicating that the suppression of $\mu_{\mathrm{DW}}$ in Ti-based trilayer is associated with a reduced wall width and modified DW structure. Such behavior is consistent with theoretical descriptions of SOT-driven DW motion, in which the flow-regime velocity depends on both the damping-like torque and the domain-wall width. Taken together, these results indicate that the variation of $\mu_{\mathrm{DW}}$ arises from the combined effects of interfacial DMI, magnetic anisotropy, and domain-wall width, demonstrating that a torque-only description is insufficient to explain the observed behavior.

These results demonstrate that although current-driven DW motion can be recovered in LM/Pt(1)/GFC trilayers, its efficiency is not governed by bulk orbital current generation in the light metal. Instead, the DW dynamics are dictated by interfacial DMI and intrinsic domain-wall structural parameters, underscoring that orbital currents alone are insufficient to control DW motion without simultaneous optimization of interfacial spin–orbit coupling and chiral stabilization.

## IV. CONCLUSIONS

In summary, we examined whether orbital Hall currents generated by current injection in Ti and Mn can directly drive DW motion in a ferrimagnetic alloy. Although finite damping-like torques are detected in Mn/GFC and Ti/GFC bilayers, their magnitudes are insufficient to induce deterministic DW motion. The small orbital torques in these systems may arise from weak orbital accumulation in Ti and Mn layers with thickness around 10 nm and/or competing orbital-to-spin conversion channels in GFC. The absence of interfacial DMI further inhibits the formation of Néel DWs, reducing the effect of the orbital torque on DW dynamics. Inserting an ultrathin Pt interlayer between the light metals and GFC simultaneously enhances orbital-to-spin conversion and induces interfacial DMI, leading to the recovery of robust DW motion. Thickness-dependent analysis reveals that the DW dynamics is governed by interfacial DMI and domain-wall width rather than by changes of orbital current generation in the light-metal layers. These results highlight that successful realization of orbital-torque-driven DW motion requires concurrent optimization of torque generation, orbital-to-spin conversion, and chiral domain-wall stabilization in orbitronic devices.

## SUPPLEMENTARY MATERIAL

The supplementary material includes additional magnetic characterization data, details of current distribution analysis, supporting domain wall measurements, discussion of the possible contribution of spin-transfer torque, and extended second-harmonic Hall data used for the quantitative evaluation of current-induced torques.

## ACKNOWLEDGMENTS

This research was supported by the Swiss National Science Foundation (Grant No. 200021_236524). M.-G.K. acknowledges support from the ETH Zürich internal funding program (Career Seed Award 24-1 SEED-008).

## AUTHOR CONTRIBUTIONS

M.-G.K and J.K. contributed equally to this work. They performed thin film growth, device fabrication, measurements, and data analysis. B.J.J. and L.V.S. provided support in the analysis of MOKE-based domain wall dynamics. M.-G.K and P.G supervised the project. The manuscript was written by M.-G.K and P.G with input from all authors.

## CONFLICT OF INTEREST

The authors declare no conflict of interests.

## DATA AVAILABILITY

The data that support the findings of this study will be made available on the ETH Research Collection repository (https://doi.org/10.3929/ethz-c-000794368).

**FIGURES**

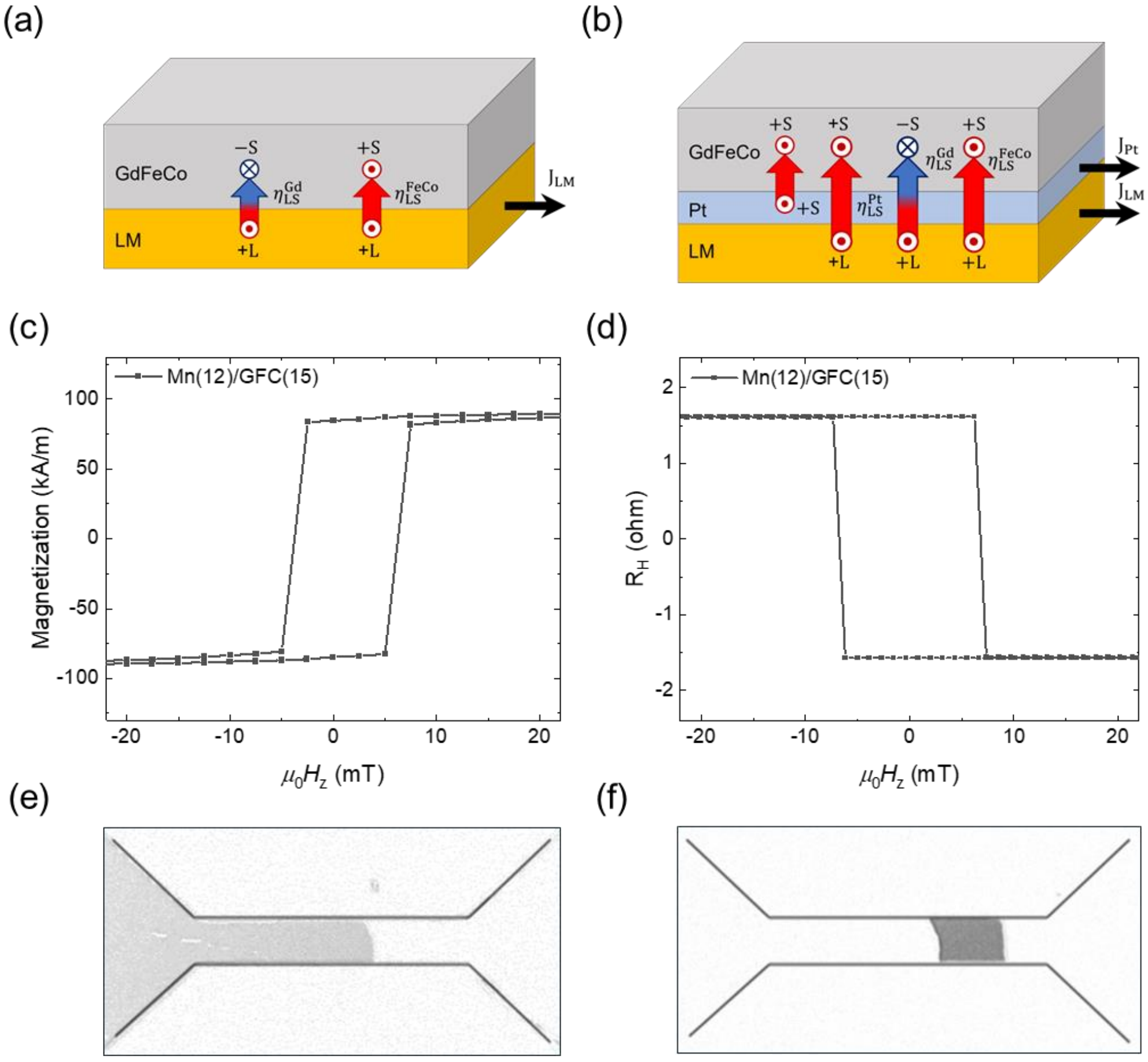

**Figure 1. (a)** Schematic of orbital-current-induced torque in LM/GFC bilayers. A charge current in the light metal generates an orbital current, which is injected into GFC and converted into spin angular momentum via L·S coupling in the Gd and FeCo sublattices. Owing to the opposite signs of spin–orbit coupling in Gd and FeCo, the converted spins have opposite polarities. **(b)** Schematic of the LM/Pt/GFC trilayer. In addition to the direct orbital-to-spin conversion inside GFC as in (a), the orbital current is converted into a spin current in Pt, and Pt also generates a spin current via the spin Hall effect, both exerting torques on GFC. **(c)** Out-of-plane hysteresis loop of a Mn(12 nm)/GFC(15 nm) sample measured by SQUID magnetometry. **(d)** Anomalous Hall resistance of Mn(12 nm)/GFC(15 nm) measured using a magnetotransport setup. The external magnetic field is applied along the out-of-plane direction. **(e)** Differential Kerr image showing nucleation of a new $+M_z$ domain from the left pad and its propagation into the racetrack. The DW is located inside the racetrack between up ($+M_z$) and down ($-M_z$) domains. **(f)** Differential Kerr image showing DW displacement driven by a series of current pulses ($t_p$ = 20 ns, $V_p$= +45 V, 10 pulses). In **(e,f)**, differential Kerr imaging captures only changes of magnetic domains by subtracting an initial frame from the subsequent image. The differential contrast in the image has been digitally enhanced to highlight the DW displacement.

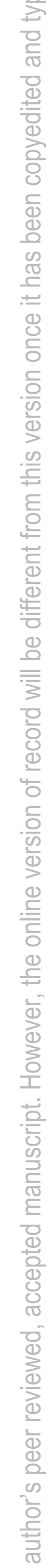

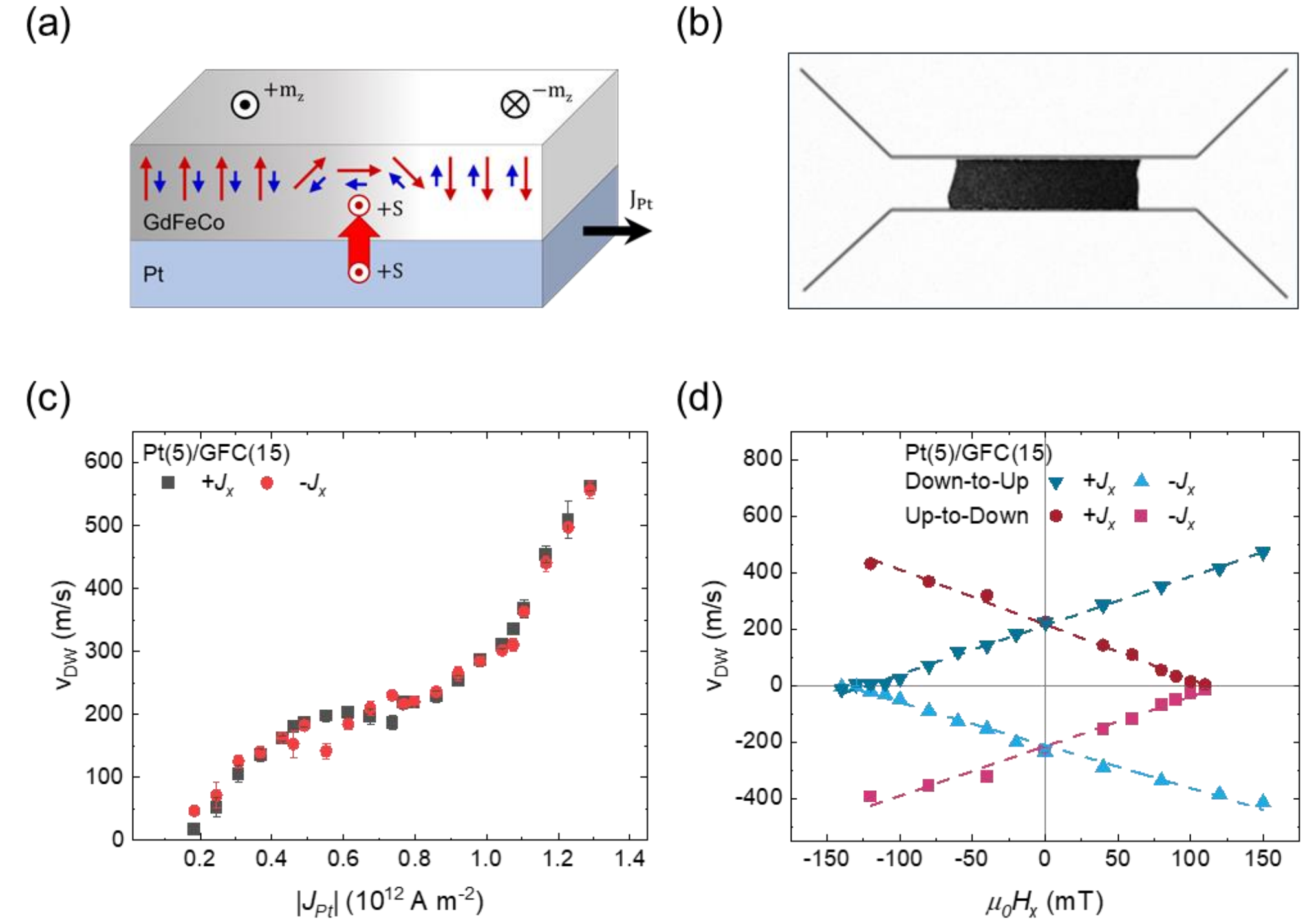

**Figure 2. (a)** Schematics of the Pt(5)/GFC(15) device structure. The sublattice magnetic moments in GFC are color-coded (Gd: red, FeCo: blue). A spin current is generated from the Pt(5) layer and injected to the adjacent GFC layer. **(b)** Differential Kerr image of SOT-driven DW displacement in Pt(5)/GFC(15) racetrack. The dark parallelogram-shaped area corresponds to the DW displacement observed by MOKE. **(c)** DW velocity $v_{DW}$ as a function of $|J_{Pt}|$ in Pt(5)/GFC(15). $+J_x$ and $-J_x$ indicate the current polarity. **(d)** $v_{DW}$ as a function of in-plane field $\mu_0 H_x$ for Down-to-Up and Up-to-Down DW configurations. The measurements were performed at $J_x = 0.92 \times 10^{12}$ A/m$^2$ and $t_p = 5$ ns.

**Figure 3. (a)** Diagram showing the effect of pulse injection in a Mn(12)/GFC(15) racetrack as a function of pulse length and amplitude. Crosses denote no domain nucleation, and red dots indicate random nucleation due to Joule heating. **(b)** Random nucleation events observed along the racetrack by MOKE. Dark contrast regions correspond to newly formed thermally nucleated domains. **(c,d)** DW displacement normalized by pulse length as a function of in-plane magnetic field $H_x$ for **(c)** Mn(12)/GFC(15) and **(d)** Mn(5)/GFC(15), measured in the Down-to-Up DW configuration. Pulse conditions are **(c)** $J_{LM} = 0.23 \times 10^{12}$ A/m$^2$, $t_p$ =30 ns and **(d)** $J_{LM} = 0.32 \times 10^{12}$ A/m$^2$, $t_p$ = 40 ns.

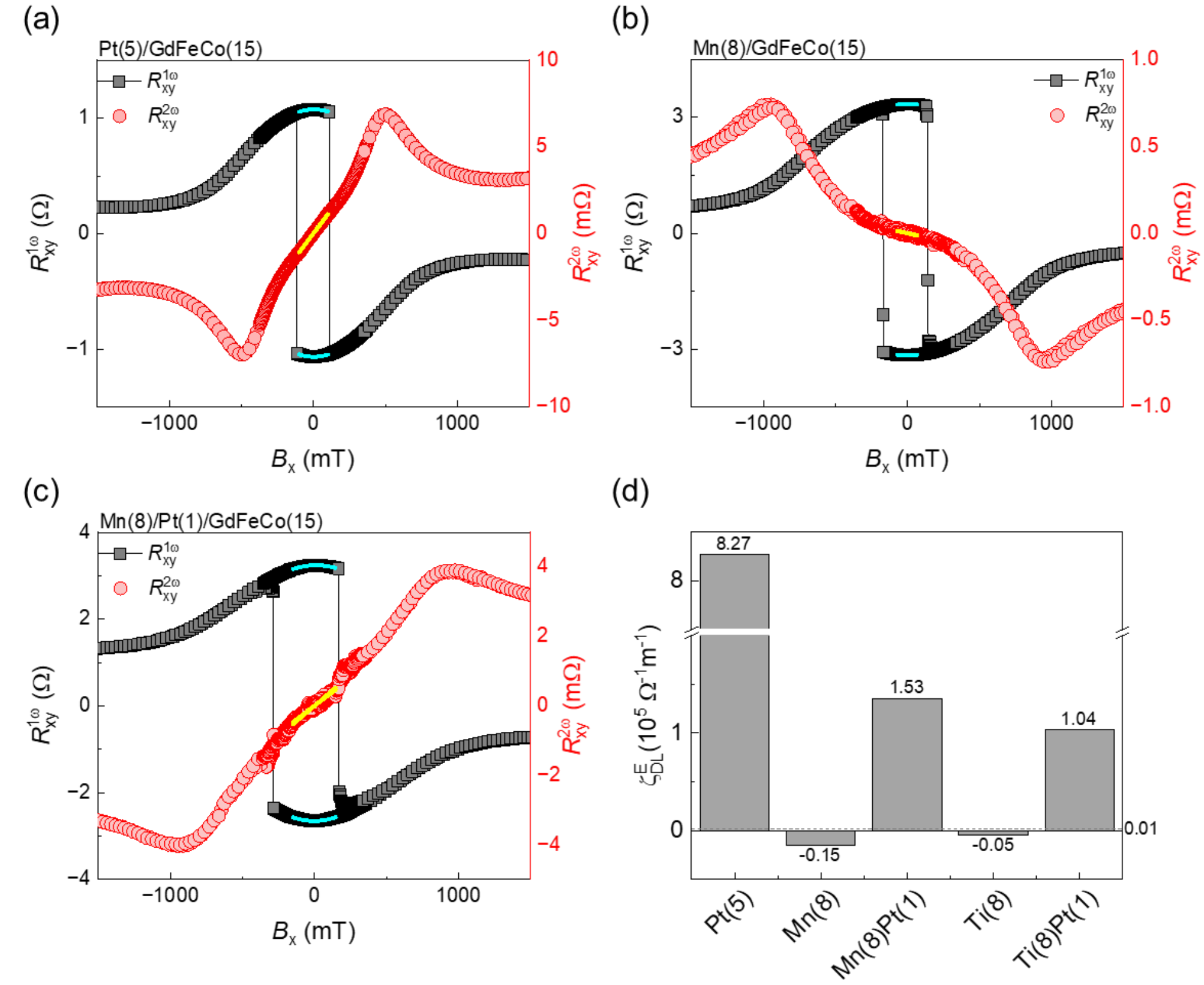

**Figure 4. (a-c)** Representative first- ($R_{xy}^{1\omega}$, black) and second-harmonic ($R_{xy}^{2\omega}$, red) Hall resistance measured as a function of in-plane magnetic field $B_x$ in **(a)** Pt(5)/GFC(15), **(b)** Mn(8)/GFC(15), and **(c)** Mn(8)/Pt(1)/GFC(15) device. Solid lines indicate parabolic fitting of $R_{xy}^{1\omega}$ (cyan) and linear fitting of $R_{xy}^{2\omega}$ (yellow) in the low-field region. **(d)** Damping-like Hall conductivity $\xi_{DL}^{E}$ extracted from second-harmonic Hall measurements for Pt(5)/GFC, Mn(8)/GFC, Mn(8)/Pt(1)/GFC, Ti(8)/GFC, and Ti(8)/Pt(1)/GFC. The gray dashed line represents the value measured in a single GFC layer.

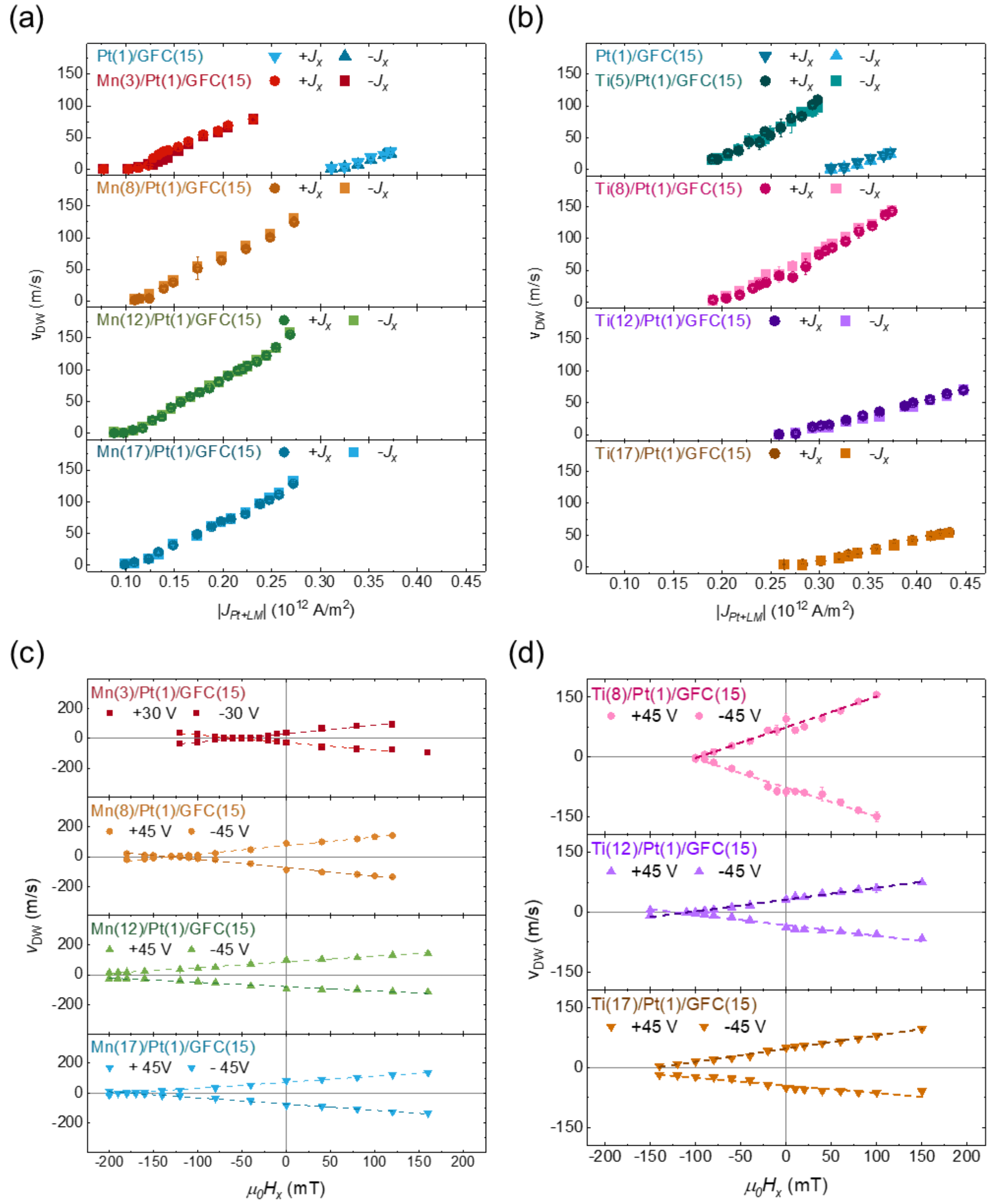

**Figure 5. (a,b)** DW velocity as a function of current density $|J_{Pt+LM}|$ for **(a)** Mn(t)/Pt(1)/GFC(15) and **(b)** Ti(t)/Pt(1)/GFC(15) trilayers. The LM thicknesses are t = 3, 8, 12, and 17 nm for Mn and t = 5, 8, 12, and 17 nm for Ti. Positive and negative current pulses drive the DW along +x and −x directions, respectively, confirming torque-driven dynamics. **(c,d)** $v_{DW}$ as a function of in-plane magnetic field $\mu_0 H_x$ for **(c)** Mn(t)/Pt(1)/GFC(15) and **(d)** Ti(t)/Pt(1)/GFC(15) devices. All measurements are performed in the Down-to-Up DW configuration with $t_p$ = 20 ns. Error bars represent the standard deviation over four measurements.

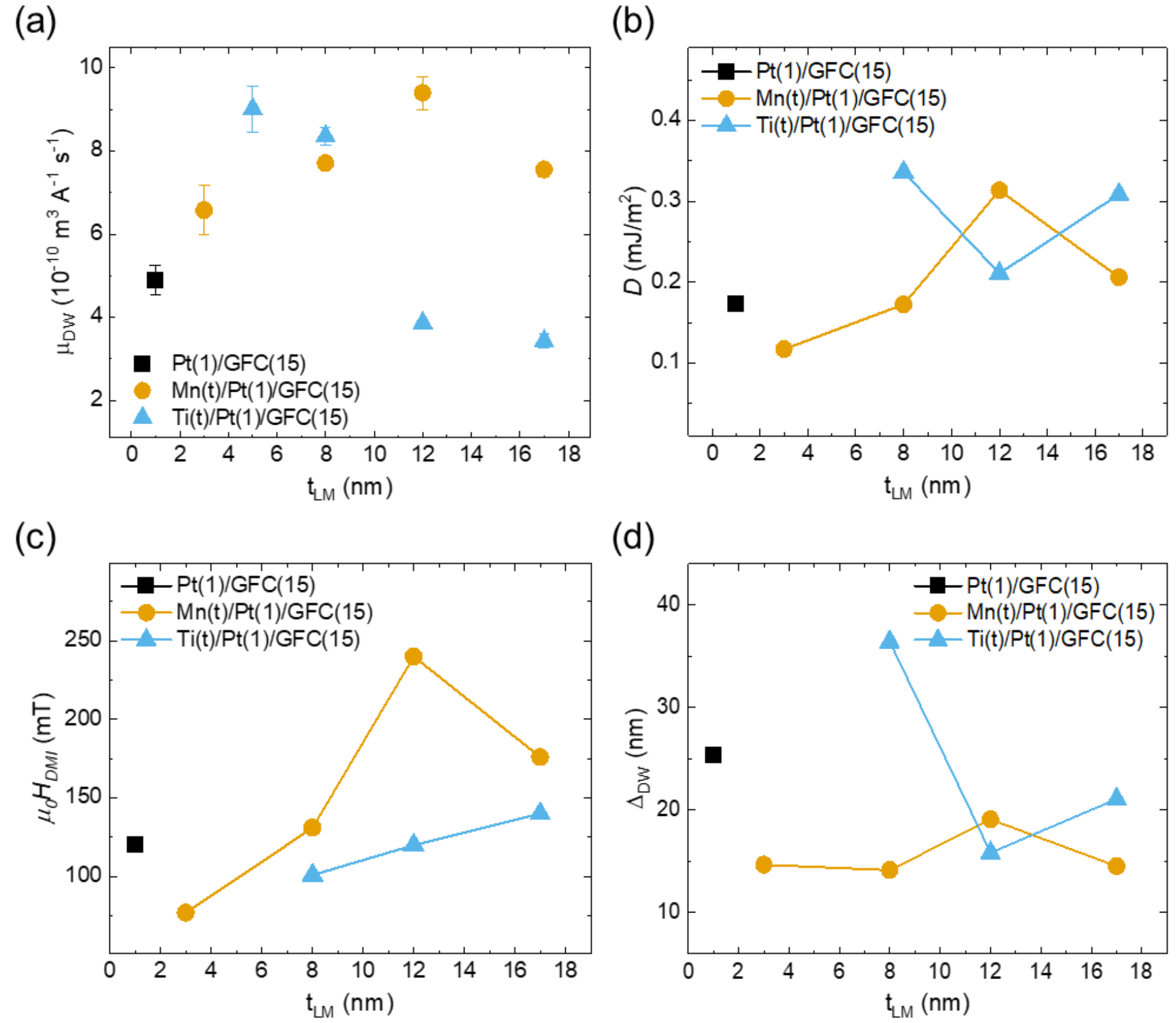

**Figure 6. (a-d) (a)** Domain wall mobility $\mu_{DW}$, **(b)** Effective DMI constant $D$, **(c)** effective DMI field $\mu_0 H_{DMI}$, and **(d)** DW width $\Delta_{DW}$ as a function of light metal thickness $t_{\mathrm{LM}}$ for Mn/Pt(1)/GFC and Ti/Pt(1)/GFC devices.